\documentclass[twocolumn,superscriptaddress,english]{revtex4}
\usepackage[T1]{fontenc}
\usepackage[latin9]{inputenc}
\usepackage{graphicx}
\usepackage{babel}
\begin{document}

\title{Interface controlled electronic variatons in correlated heterostructures}
\author{K. Gehrke}
\author{V. Moshnyaga}
\author{K. Samwer}
\affiliation{I. Physikalisches Institut, Georg-August-Universität Göttingen, Friedrich-Hund-Platz 1, 37077 Göttingen, Germany}
\author{O. I. Lebedev}
\author{D.Kirilenko}
\author{G.Van Tendeloo}
\affiliation{EMAT, University of Antwerp, Groenenborgerlaan 171, B-2020 Antwerpen, Belgium }

\begin{abstract}
An interface modification of $\mathrm{(LaCa)MnO_{3}-BaTiO_{3}}$ superlattices was found to massively influence magnetic and magnetotransport properties. Moreover it determines the crystal structure of the manganite layers, changing it from orthorhombic (Pnma) for the conventional superlattice (cSL), to rhombohedral $\mathrm{(R\bar{3}c)}$ for the modified one (mSL). While the cSL shows extremely nonlinear ac transport, the mSL is an electrically homogeneous material. The observations go beyond an oversimplified picture of dead interface layers and evidence the importance of electronic correlations at perovskite interfaces.
\end{abstract}

\maketitle

Interfaces of complex oxides have gained much attention since the
discovery of a high-mobility quasi two dimensional electron gas at
the $\mathrm{TiO_{2}/LaO}$-interface between insulating $\mathrm{LaAlO_{3}}$
(LAO) and $\mathrm{SrTiO_{3}}$ (STO) \citep{Ohtomo2004,Thiel2006}.
This unexpected finding disclosed a new role of interfaces in oxide
heterostructures. Thereby the reconstruction of the interface is discussed
to avoid an electrostatic potential, otherwise building up in the
lanthanum perovskite. For this reconstruction either electrons redistribute
or ions rearrange through relaxation or deviation from stoichiometry.
The charge transfer can effectively dope the materials in a rather
thin region in the vicinity of the interface \citep{Dagotto2009,Nakagawa2006},
and the resulting change in carrier concentration due to the presence
of an interface can be termed {}``interface doping''. For strongly
correlated electron systems it is well known that the carrier concentration
massively influences the properties of the material. A prominent example
is the class of perovskite manganites, which are interesting not only
because of high spin polarization in the ferromagnetic phase \citep{Muller2009,Nadgorny2001},
but also due to the fascinating rich magnetic phase diagram that opens
up upon doping \citep{Schiffer1995}. The ground state of $\mathrm{La_{1-x}Ca_{x}MnO_{3}}$
(LCMO) changes from ferromagnetic (FM) metallic for $0.2\le x\le0.4$
to antiferromagnetic (AFM) insulating for $x\ge0.5$, whereas the
phase boundary is not a sharp line, but rather a broad region around
$x=0.45$ where FM and AFM phases coexist \citep{Brink1999}. The
FM state is stabilized by gaining kinetic energy due to the delocalization
of charge carriers at the expense of antiferromagnetic exchange of
localized spins. Localization is stabilized by the Jahn-Teller (JT)
effect, that lifts the $\mathrm{e_{g}}$-orbital degeneracy of $\mathrm{Mn^{3+}}$
ions \citep{Khomskii2001}, giving rise to JT-polarons, that are discussed
to be the main type of charge carrier at all temperatures \citep{Zhao2000}.
Binding of these JT-polarons into pairs of correlated polarons (CP)
or bipolarons is now argued to bring about the strong localization
at the metal insulator (MI) transition \citep{Alexandrov1999}. Even
for optimal doping $(0.2\le x\le0.4)$ correlated JT polarons have
been observed by neutron and x-ray scattering \citep{Adams2000,Nelson2001},
with the wave vector of these short range JT-distorted regions being
$\vec{q}=[\frac{1}{4},\,\frac{1}{4},\,0]$.

The loss of FM order in $\mathrm{La_{\frac{2}{3}}Ca_{\frac{1}{3}}MnO_{3}}$
(LCMO) and $\mathrm{La_{\frac{2}{3}}Sr_{\frac{1}{3}}MnO_{3}}$ (LSMO) below a critical film thickness \citep{Bibes2002,Huijben2008},
and the poor performance of LSMO-STO-LSMO tunnel junctions at elevated temperatures \citep{Yamada2004,Ishii2006},
have been connected to the weakening of double exchange at manganite
interfaces. The breakup of the Mn-O chains has been accounted for this interface-induced phase separation
\citep{Giesen2004} and the carrier depletion (interface doping),
alike the one discussed for LAO/STO-interfaces, seems to be a driving
force of the localization. Moreover it has been shown that two monolayers
of $\mathrm{LaMnO_{3}}$ (LMO) introduce extra carriers to the interface
region, counteracting the depletion and stabilizing the FM/spin polarization
\citep{Yamada2004,Ishii2006,Brey2007}. Elastic constraints also play
an important role, since lattice relaxations are intrinsically coupled
to the electronic correlations in Jahn-Teller Systems \citep{Leonov2008}.
Furthermore CP can be attributed to the CE-AFM phase, which shows the same superstructure wave vector
$\vec{q}=[\frac{1}{4},\,\frac{1}{4},\,0]$. As we have shown recently,
CP can be probed by means of the 3rd harmonic voltage, which is a
measure of electric nonlinearity, because of their quadrupole nature
\citep{Moshnyaga2009}. Here we use AC-transport, SQUID magnetometry and transmission electron
microscopy (TEM) to study the influence of interfaces in SLs and the
effect of an interface modification by additional LMO layers.

We compare the physical properties of two manganite-titanate superlattices
(SL). The conventional SL (cSL) consists of 40 unit cells (u.c.) LCMO
and 20 u.c. $\mathrm{BaTiO_{3}}$ (BTO), repeated ten times $\mathrm{[LCMO_{40}/BTO_{20}]_{10}}$.
The modified SL (mSL) consists of the same LCMO and BTO layers, but
additionally two u.c. of LMO were introduced at each interface: $\mathrm{[LCMO_{40}/LMO_{2}/BTO_{20}/LMO_{2}]_{10}}$.
The SLs were grown on MgO (100) substrates at $T=900\,^{\circ}\mathrm{C}$ by
means of a metal organic aerosol deposition technique \citep{Moshnyaga1999}.
The mono-layer accuracy during deposition was achieved by controlling
the volume of the metal organic solution for each layer.

\begin{figure}
\includegraphics[width=7cm]{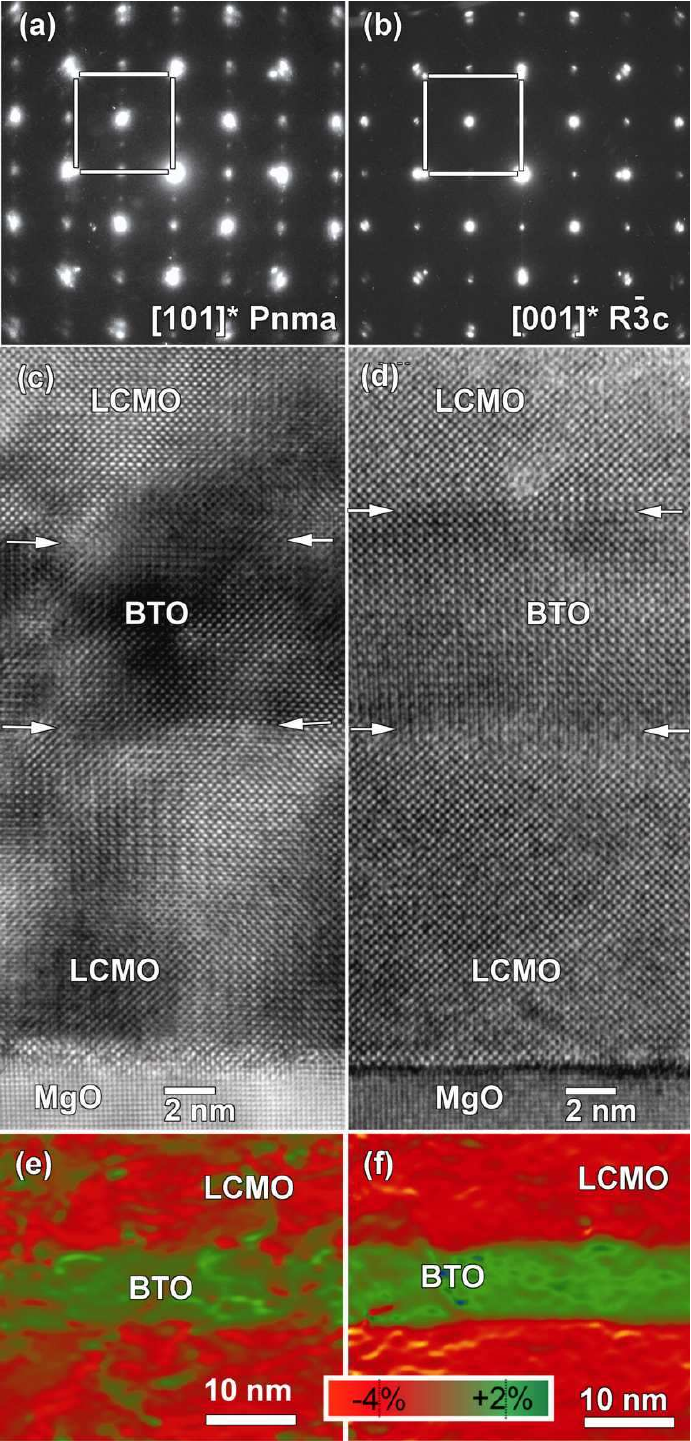}\caption{\label{fig:TEM}Microstructure analysis of the cSL (left) and mSL
(right). Electron diffraction patterns (top) reveal Pnma-symmetry
for cSL (a) and $\mathrm{R\bar{3}c}$-symmetry for mSL (b). Cross-section
HRTEM (mid) confirm epitaxial growth and coherent interfaces of cSL
(c) and mSL (d). GPA (bottom) shows strain fields along the growth
direction to be inhomogeneous for cSL (e) and homogeneous for mSL
(f)}
\end{figure}

The XRD ($\Theta$-$2\Theta$) patterns show typical satellite peaks,
arising from the SL periodicity (not shown). The calculated periodicities
from Schuller's formula \citep{Schuller1980} are $\lambda_{cSL}=23.3\pm0.05\, nm$
and $\lambda_{mSL}=25.3\pm0.05\, nm$, which is in a good agreement
with the nominal thicknesses. A detailed study of the crystal and
chemical structure of SLs was performed on cross-section and plan
view specimens by means of TEM.
Plan view and cross section electron diffraction (ED) (Fig.\ref{fig:TEM}a,b)
patterns clearly show heteroepitaxial growth of all layers for both
SLs. Cross-section high resolution TEM (HRTEM) images confirm epitaxial
growth of SLs and show coherent, atomically flat and sharp interfaces
(Fig.1c,d). The extra LMO layers in mSL are not visible in HRTEM image
but, obviously, introduce significant structural differences between
cSL and mSL as it can be detected by ED and HRTEM. The LCMO layer
of the mSL exhibits the unusual $\mathrm{R\bar{3}c}$ structure in
contrast to most bulk samples \citep{Radaelli1996} and cSL LCMO showing
both Pnma structure. The structure of BTO of both SLs is found to
be close to P4mm. The tetragonality of the BTO-layers is $\frac{c}{a}=1.04\pm0.005$
for the conventional and $\frac{c}{a}=1.025\pm0.005$ for the modified
SL. Energy filtered TEM (EFTEM) and electron energy loss spectroscopy
(EELS) measurements of SLs (not shown) show well defined chemical
separation of LCMO and BTO layers leading to the conclusion that there
is no large interdiffusion and the interfaces are chemically sharp.
In order to analyze strain fields, the geometrical phase analysis
(GPA) method \citep{Hytch1995} has been applied to HRTEM images of
cSL and mSL. The results of GPA along the growth direction are shown
in Fig.\ref{fig:TEM}(e), (f) and reveal inhomogeneously strained
cSL, in contrast to the mSL heterostructure which looks homogeneously
strained. %

\begin{figure}
\includegraphics[width=8cm]{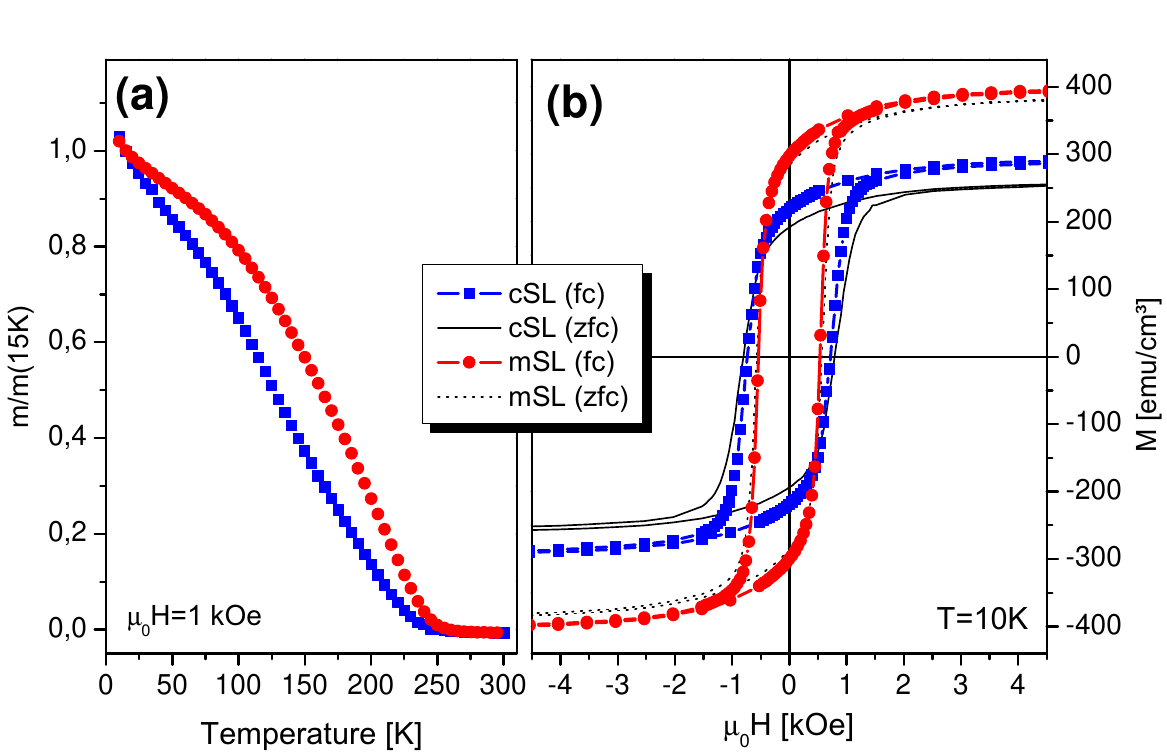}\caption{\label{fig:M(H)@10K}(a) Normalized magnetization as a function of
temperature of conventional (cSL) and modified (mSL) superlattice
samples. (b) Magnetic Hysteresis of cSL and mSL at $T=10\, K$. Measurements
taken after field cooling (fc) and zero field cooling (zfc) are shown.}

\end{figure}

Magnetic properties of the SLs have been measured with external field applied parallel to the
film plane. The temperature dependence of the magnetization of both
SLs, see Fig.\ref{fig:M(H)@10K}(a) shows a rather broad ferromagnetic
transition, typical for thin LCMO-layers \citep{Gong1996,Jo1999}.
The magnetic Curie temperatures of the SLs are $T_{C}^{cSL}=245\, K$
and $T_{C}^{mSL}=253\, K$. Magnetic hysteresis M(H), shown in Fig.\ref{fig:M(H)@10K}(b)
was measured at T=10K after cooling without (zfc) and
with an external field $\mu_{0}H=5\, T$ applied during cooling (fc).
The saturation magnetization $(M_{s})$ of the SLs after fc is as
large as $M_{s}^{cSL}=287\,\frac{emu}{cm^{3}}$ and $M_{s}^{mSL}=393\,\frac{emu}{cm^{3}}$,
which is smaller than the bulk value, $M_{s}^{bulk}\approx650\,\frac{emu}{cm^{3}}$.
In the case of the mSL, additional LMO layers are expected to contribute
to the magnetic moment and therefore the volumes of both LCMO and
LMO were considered calculating $M_{s}^{mSL}$. The corresponding
coercivities are $H_{c}^{cSL}=721\, Oe$ and $H_{c}^{mSL}=536\, Oe$.
Interestingly, after zfc the $M_{s}$ of both SLs were smaller than
$M_{s}$ after fc, $\frac{\Delta M_{s}}{M_{s}}\approx-11\%$ for the
cSL and $\frac{\Delta M_{s}}{M_{s}}\approx-3\%$ for the mSL.%

\begin{figure}
\includegraphics[width=8cm]{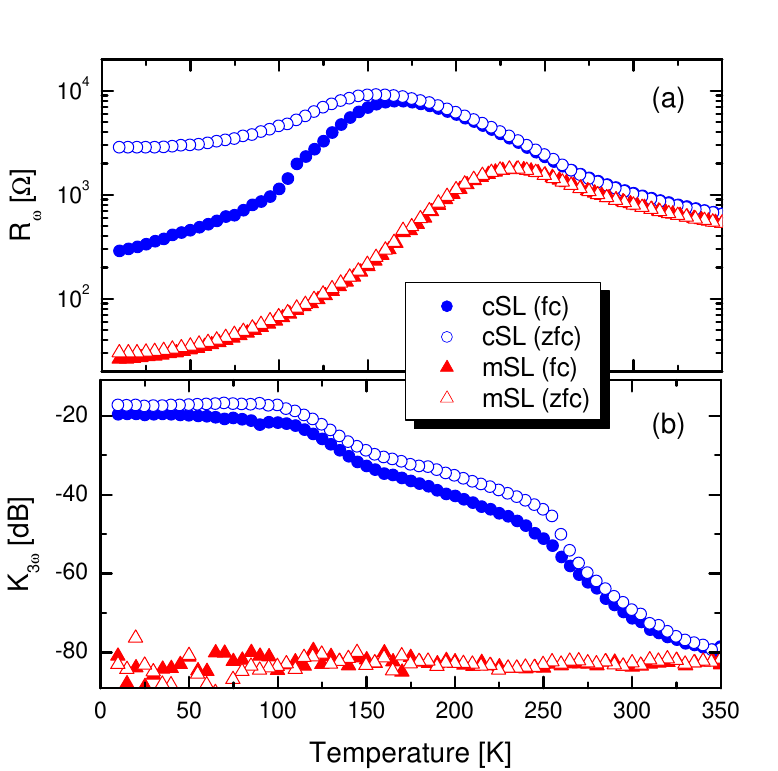}\caption{\label{fig:SL-transport}Transport properties of the conventional
(cSL) and modified (mSL) superlattice as a function of temperature.
Linear ac-resistance $R_{\omega}$(a) and the nonlinear coefficient
$K_{3\omega}$ (b) are shown for field cooled (fc) and zero field
cooled (zfc) measurements.}

\end{figure}

Measurements of the ac electric transport were performed by means
of four probe method, with silver paste contacts at the edges of the
SLs. The ac-current amplitude was $I_{ac}=10\,µA$ and the frequency
$f_{ac}=17\, Hz$. Besides the linear electric response with resistance
$R_{\omega}$ at the fundamental frequency $f_{ac}$, we report the
third harmonic response at $f_{3\omega}=3f_{ac}$ in terms of the
coefficient $K_{3\omega}=\frac{U_{3\omega}}{U_{\omega}}$, which is
a measure of the electrical nonlinearity in the film \citep{Moshnyaga2009}.
Fig.\ref{fig:SL-transport}(a) shows the temperature dependence of
$R_{\omega}$, measured after field cooling (fc) and zero field cooling
(zfc). The cSL shows a large difference (factor of ten) between fc
and zfc resistance at low temperature, $T=10\, K$. The difference
gets smaller at $T\approx150\, K$ and vanishes completely close to
$T_{c}=245\, K$. The metal-insulator (MI) transition of the cSL is
shifted by $\Delta T=77\, K$ below the magnetic transition. For the
mSL no difference between $R_{\omega}^{mSL}(fc)$ and $R_{\omega}^{mSL}(zfc)$
was observed, and $T_{c}^{mSL}-T_{MI}^{mSL}=22\, K$. Regarding nonlinear
transport, shown in Fig.\ref{fig:SL-transport}(b), the cSL shows
a strong increase of $K_{3\omega}$by decreasing temperature, both
for fc- and zfc-condition. Two features in $K_{3\omega}(T)$ are distinguishable:
at $T=100\, K$ and $T=255\, K$ there exist kinks. Below $T=100\, K$
the non linearity is very large, $K_{3\omega}\approx-20\, dB$, and
nearly constant. For the mSL a very small nonlinear signal $(K_{3\omega}<-80\, dB)$
was measured in the whole temperature range, both for fc- and zfc-condition.%

In the case of manganite-titanate interfaces, the concept of interface
doping leads to an overall increase of the Mn-valence, which would
be equal to an increase of doping to $x=x_{nominal}+x_{interface}$.
Recent theoretical and experimental studies show that the length scale
on which charge transfer takes place is rather small, $l_{ct}\approx1\, nm$
\citep{Gonzalez2008,Adamo2009}, and therefore $x$ is increased only
in the vicinity of the interfaces. By adding LMO-layers we decrease
$x_{nominal}$ in this region. The overall doping level at the interfaces
of the mSL is smaller compared to that at the cSL interfaces. In this
sense, the interface regions bring about different electronic constraints
to the LCMO-layers in the SLs. Because of the strong electronic correlations
these electronic constraints have tremendous impact on the structural
properties of the LCMO-layers.

First of all the enhanced saturation magnetization $M_{s}$ and reduced
coercivity $H_{c}$ in mSL, see Fig.\ref{fig:M(H)@10K}(a), directly
show that the interface modification strongly reduces the tendency
towards AFM correlation. The difference in $M_{s}$ after fc/zfc and
the extreme difference between $T_{C}$ and $T_{MI}$ for the cSL
underlines a pronounced magnetic inhomogeneity of the cSL. Due to
the modification magnetic inhomogeneity is reduced in the mSL, which
can also be seen in a somewhat steeper magnetic transition (see Fig.\ref{fig:M(H)@10K}(b)).
The difference of the magnetic homogeneity is also reflected by the
electric transport behavior of the SLs. The fc/zfc-difference, observed
only for the cSL (see Fig.\ref{fig:SL-transport}(a)), can be attributed
to an inhomogeneous electronic (phase separated) state, which is known
from $\mathrm{La_{1-x}Ca_{x}MnO_{3}}$ with $0.4\le x\le0.5$.

Most interestingly, the different electric constraints of the SLs
not only change the magnetic and transport properties, but also the
structure of the interface region and hence of the whole LCMO-layers.
For LCMO without oxygen deficiency, it was shown that crystalline
structure, determined by the amount of $\mathrm{Mn^{4+}}$-ions, changes
from $\mathrm{R\bar{3}c}$ for $x\le0.4$ to Pnma for $x\ge0.5$ \citep{Yahia2003}.
The right oxygenation can be expected in the case of metal organic
aerosol deposition technique, as films are grown at ambient oxygen
pressure. The manganite of the mSL shows $\mathrm{R\bar{3}c}$ symmetry,
which has been observed for small amounts of $\mathrm{Mn^{4+}}$-Ions,
i.e. $x\le0.4$. The LCMO-layers of the cSL show Pnma-symmetry, which
is found for $x\ge0.5$. So, most probably the interface region of
cSL is in the CE-AFM state, as it is found for LCMO with $x\ge0.5$.
As we have shown previously \citep{Moshnyaga2009} the manganites
with $\mathrm{R\bar{3}c}$ structure (LSMO) show linear transport
behavior, whereas for the Pnma structure non linearity has been observed
in the transition region. Here we see another example of this rule:
the mSL $\mathrm{(R\bar{3}c)}$ shows a very small $K_{3\omega}\approx-80\, dB$,
whereas the cSL (Pnma) shows pronounced nonlinear transport properties,
see Fig.\ref{fig:SL-transport}(b). The extremely large nonlinear
signal, $K_{3\omega}>-20\, dB$, at low temperatures on the other
hand is in contrast to the results for single LCMO-films \citep{Moshnyaga2009}.
This can be attributed to the high density of interfaces of the cSL
compared to a single LCMO-film. The interfaces act like nucleation
centers, at which the CP can accumulate and therefore the amount of
CP is enhanced in cSL. If one assumes the CP to be the building blocks
of the CE phase, it is clear that the interface region is unstable
against the antiferromagnetic CE-correlations \citep{Brey2007}. For
the mSL the number of CP is reduced, so that $K_{3\omega}$ is very
small. Furthermore the Pnma-symmetry, in contrast to the homogeneous
$\mathrm{R\bar{3}c}$-symmetry, allows for local elastic deformation.
Phase separation into FM and CE-AFM, like it has been discussed for
LCMO with $0.4\le x\le0.5$ \citep{Brink1999}, can lead to the inhomogeneous
magnetic and electric state of the cSL. 

In summary we have shown that engineering of the doping profile in
LCMO-BTO superlattices has massive influences on the magnetotransport
properties and structure of the manganite layers. Only two monolayers
of LMO at each interface lead to an enhancement of the saturation
magnetization. We discussed, that in conventional LCMO/BTO superlattices
the ground state of the interfacial LCMO layers is likely CE-AFM,
supporting theoretical calculations \citep{Brey2007}. This phase
is responsible for exceptional nonlinear $(3\omega)$ transport as
well as for zfc-fc splitting of the linear resistance discussed within
correlated polaron model and phase separation scenario. This work
was supported by DFG via SFB 602, TPA2.


\begin{thebibliography}{10}
\providecommand*{\bibinfo}[2]{#2}
\providecommand*{\eprint}[1]{#1}
\providecommand*{\url}[1]{#1}
\bibitem{Ohtomo2004}
\bibinfo{author}{A.~Ohtomo} and \bibinfo{author}{H.~Y. Hwang},
  \bibinfo{journal}{Nature} \bibinfo{volume}{\textbf{427}}(6973),
  \bibinfo{pages}{423} (\bibinfo{date}{Jan. 2004}).
\bibitem{Thiel2006}
\bibinfo{author}{S.~Thiel}, \bibinfo{author}{G.~Hammerl},
  \bibinfo{author}{A.~Schmehl}, \bibinfo{author}{C.~W. Schneider}, and
  \bibinfo{author}{J.~Mannhart}, \bibinfo{journal}{Science}
  \bibinfo{volume}{\textbf{313}}(5795), \bibinfo{pages}{1942}
  (\bibinfo{date}{2006}).
\bibitem{Dagotto2009}
\bibinfo{author}{E.~Dagotto}, \bibinfo{journal}{Physics}
  \bibinfo{volume}{\textbf{2}}, \bibinfo{pages}{12}, \bibinfo{eid}{12}
  (\bibinfo{date}{Feb 2009}).
\bibitem{Nakagawa2006}
\bibinfo{author}{N.~Nakagawa}, \bibinfo{author}{H.~Y. Hwang}, and
  \bibinfo{author}{D.~A. Muller}, \bibinfo{journal}{Nat Mater}
  \bibinfo{volume}{\textbf{5}}(3), \bibinfo{pages}{204} (\bibinfo{date}{Mar.
  2006}).
\bibitem{Muller2009}
\bibinfo{author}{G.~M. Muller}, \bibinfo{author}{J.~Walowski},
  \bibinfo{author}{M.~Djordjevic}, \bibinfo{author}{G.-X. Miao},
  \bibinfo{author}{A.~Gupta}, \bibinfo{author}{A.~V. Ramos},
  \bibinfo{author}{K.~Gehrke}, \bibinfo{author}{V.~Moshnyaga},
  \bibinfo{author}{K.~Samwer}, \bibinfo{author}{J.~Schmalhorst}, \emph{et~al.},
  \bibinfo{journal}{Nat Mater} \bibinfo{volume}{\textbf{8}}(1),
  \bibinfo{pages}{56} (\bibinfo{date}{Jan. 2009}).
\bibitem{Nadgorny2001}
\bibinfo{author}{B.~Nadgorny}, \bibinfo{author}{I.~I. Mazin},
  \bibinfo{author}{M.~Osofsky}, \bibinfo{author}{R.~J. Soulen},
  \bibinfo{author}{P.~Broussard}, \bibinfo{author}{R.~M. Stroud},
  \bibinfo{author}{D.~J. Singh}, \bibinfo{author}{V.~G. Harris},
  \bibinfo{author}{A.~Arsenov}, and \bibinfo{author}{Y.~Mukovskii},
  \bibinfo{journal}{Phys. Rev. B} \bibinfo{volume}{\textbf{63}}(18),
  \bibinfo{pages}{184433} (\bibinfo{date}{Apr. 2001}).
\bibitem{Schiffer1995}
\bibinfo{author}{P.~Schiffer}, \bibinfo{author}{A.~P. Ramirez},
  \bibinfo{author}{W.~Bao}, and \bibinfo{author}{S.-W. Cheong},
  \bibinfo{journal}{Phys. Rev. Lett.} \bibinfo{volume}{\textbf{75}}(18),
  \bibinfo{pages}{3336} (\bibinfo{date}{Oct. 1995}).
\bibitem{Brink1999}
\bibinfo{author}{J.~van~den Brink}, \bibinfo{author}{G.~Khaliullin}, and
  \bibinfo{author}{D.~Khomskii}, \bibinfo{journal}{Phys. Rev. Lett.}
  \bibinfo{volume}{\textbf{83}}(24), \bibinfo{pages}{5118} (\bibinfo{date}{Dec
  1999}).
\bibitem{Khomskii2001}
\bibinfo{author}{D.~I. Khomskii}, \bibinfo{journal}{International Journal of
  Modern Physics B} \bibinfo{volume}{\textbf{15}}, \bibinfo{pages}{2665}
  (\bibinfo{date}{2001}).
\bibitem{Zhao2000}
\bibinfo{author}{G.-m. Zhao}, \bibinfo{author}{V.~Smolyaninova},
  \bibinfo{author}{W.~Prellier}, and \bibinfo{author}{H.~Keller},
  \bibinfo{journal}{Phys. Rev. Lett.} \bibinfo{volume}{\textbf{84}}(26),
  \bibinfo{pages}{6086} (\bibinfo{date}{Jun. 2000}).
\bibitem{Alexandrov1999}
\bibinfo{author}{A.~S. Alexandrov} and \bibinfo{author}{A.~M. Bratkovsky},
  \bibinfo{journal}{Phys. Rev. Lett.} \bibinfo{volume}{\textbf{82}}(1),
  \bibinfo{pages}{141} (\bibinfo{date}{Jan. 1999}).
\bibitem{Adams2000}
\bibinfo{author}{C.~P. Adams}, \bibinfo{author}{J.~W. Lynn},
  \bibinfo{author}{Y.~M. Mukovskii}, \bibinfo{author}{A.~A. Arsenov}, and
  \bibinfo{author}{D.~A. Shulyatev}, \bibinfo{journal}{Phys. Rev. Lett.}
  \bibinfo{volume}{\textbf{85}}(18), \bibinfo{pages}{3954} (\bibinfo{date}{Oct.
  2000}).
\bibitem{Nelson2001}
\bibinfo{author}{C.~S. Nelson}, \bibinfo{author}{M.~v. Zimmermann},
  \bibinfo{author}{Y.~J. Kim}, \bibinfo{author}{J.~P. Hill},
  \bibinfo{author}{D.~Gibbs}, \bibinfo{author}{V.~Kiryukhin},
  \bibinfo{author}{T.~Y. Koo}, \bibinfo{author}{S.-W. Cheong},
  \bibinfo{author}{D.~Casa}, \bibinfo{author}{B.~Keimer}, \emph{et~al.},
  \bibinfo{journal}{Phys. Rev. B} \bibinfo{volume}{\textbf{64}}(17),
  \bibinfo{pages}{174405} (\bibinfo{date}{Oct. 2001}).
\bibitem{Bibes2002}
\bibinfo{author}{M.~Bibes}, \bibinfo{author}{S.~Valencia},
  \bibinfo{author}{L.~Balcells}, \bibinfo{author}{B.~Mart\'inez},
  \bibinfo{author}{J.~Fontcuberta}, \bibinfo{author}{M.~Wojcik},
  \bibinfo{author}{S.~Nadolski}, and \bibinfo{author}{E.~Jedryka},
  \bibinfo{journal}{Phys. Rev. B} \bibinfo{volume}{\textbf{66}}(13),
  \bibinfo{pages}{134416} (\bibinfo{date}{Oct 2002}).
\bibitem{Huijben2008}
\bibinfo{author}{M.~Huijben}, \bibinfo{author}{L.~W. Martin},
  \bibinfo{author}{Y.-H. Chu}, \bibinfo{author}{M.~B. Holcomb},
  \bibinfo{author}{P.~Yu}, \bibinfo{author}{G.~Rijnders},
  \bibinfo{author}{D.~H.~A. Blank}, and \bibinfo{author}{R.~Ramesh},
  \bibinfo{journal}{Physical Review B (Condensed Matter and Materials Physics)}
  \bibinfo{volume}{\textbf{78}}(9), \bibinfo{pages}{094413},
  \bibinfo{eid}{094413} (\bibinfo{numpages}{7}  pages)  (\bibinfo{date}{2008}).
\bibitem{Yamada2004}
\bibinfo{author}{H.~Yamada}, \bibinfo{author}{Y.~Ogawa},
  \bibinfo{author}{Y.~Ishii}, \bibinfo{author}{H.~Sato},
  \bibinfo{author}{M.~Kawasaki}, \bibinfo{author}{H.~Akoh}, and
  \bibinfo{author}{Y.~Tokura}, \bibinfo{journal}{Science}
  \bibinfo{volume}{\textbf{305}}(5684), \bibinfo{pages}{646}
  (\bibinfo{date}{2004}).
\bibitem{Ishii2006}
\bibinfo{author}{Y.~Ishii}, \bibinfo{author}{H.~Yamada},
  \bibinfo{author}{H.~Sato}, \bibinfo{author}{H.~Akoh},
  \bibinfo{author}{Y.~Ogawa}, \bibinfo{author}{M.~Kawasaki}, and
  \bibinfo{author}{Y.~Tokura}, \bibinfo{journal}{Applied Physics Letters}
  \bibinfo{volume}{\textbf{89}}(4), \bibinfo{pages}{042509},
  \bibinfo{eid}{042509} (\bibinfo{numpages}{3}  pages)  (\bibinfo{date}{2006}).
\bibitem{Giesen2004}
\bibinfo{author}{F.~Giesen}, \bibinfo{author}{B.~Damaschke},
  \bibinfo{author}{V.~Moshnyaga}, \bibinfo{author}{K.~Samwer}, and
  \bibinfo{author}{G.~A. M\"uller}, \bibinfo{journal}{Phys. Rev. B}
  \bibinfo{volume}{\textbf{69}}(1), \bibinfo{pages}{014421} (\bibinfo{date}{Jan
  2004}).
\bibitem{Brey2007}
\bibinfo{author}{L.~Brey}, \bibinfo{journal}{Physical Review B (Condensed
  Matter and Materials Physics)} \bibinfo{volume}{\textbf{75}}(10),
  \bibinfo{pages}{104423}, \bibinfo{eid}{104423} (\bibinfo{numpages}{7}  pages)
   (\bibinfo{date}{2007}).
\bibitem{Leonov2008}
\bibinfo{author}{I.~Leonov}, \bibinfo{author}{N.~Binggeli},
  \bibinfo{author}{D.~Korotin}, \bibinfo{author}{V.~I. Anisimov},
  \bibinfo{author}{N.~Stojic}, and \bibinfo{author}{D.~Vollhardt},
  \bibinfo{journal}{Physical Review Letters} \bibinfo{volume}{\textbf{101}}(9),
  \bibinfo{pages}{096405}, \bibinfo{eid}{096405} (\bibinfo{numpages}{4}  pages)
   (\bibinfo{date}{2008}).
\bibitem{Moshnyaga2009}
\bibinfo{author}{V.~Moshnyaga}, \bibinfo{author}{K.~Gehrke},
  \bibinfo{author}{O.~I. Lebedev}, \bibinfo{author}{L.~Sudheendra},
  \bibinfo{author}{A.~Belenchuk}, \bibinfo{author}{S.~Raabe},
  \bibinfo{author}{O.~Shapoval}, \bibinfo{author}{J.~Verbeeck},
  \bibinfo{author}{G.~V. Tendeloo}, and \bibinfo{author}{K.~Samwer},
  \bibinfo{journal}{Physical Review B (Condensed Matter and Materials Physics)}
  \bibinfo{volume}{\textbf{79}}(13), \bibinfo{pages}{134413},
  \bibinfo{eid}{134413} (\bibinfo{numpages}{8}  pages)  (\bibinfo{date}{2009}).
\bibitem{Moshnyaga1999}
\bibinfo{author}{V.~Moshnyaga}, \bibinfo{author}{I.~Khoroshun},
  \bibinfo{author}{A.~Sidorenko}, \bibinfo{author}{P.~Petrenko},
  \bibinfo{author}{A.~Weidinger}, \bibinfo{author}{M.~Zeitler},
  \bibinfo{author}{B.~Rauschenbach}, \bibinfo{author}{R.~Tidecks}, and
  \bibinfo{author}{K.~Samwer}, \bibinfo{journal}{Applied Physics Letters}
  \bibinfo{volume}{\textbf{74}}(19), \bibinfo{pages}{2842}
  (\bibinfo{date}{1999}).
\bibitem{Schuller1980}
\bibinfo{author}{I.~K. Schuller}, \bibinfo{journal}{Phys. Rev. Lett.}
  \bibinfo{volume}{\textbf{44}}(24), \bibinfo{pages}{1597} (\bibinfo{date}{Jun.
  1980}).
\bibitem{Radaelli1996}
\bibinfo{author}{P.~G. Radaelli}, \bibinfo{author}{M.~Marezio},
  \bibinfo{author}{H.~Y. Hwang}, \bibinfo{author}{S.-W. Cheong}, and
  \bibinfo{author}{B.~Batlogg}, \bibinfo{journal}{Phys. Rev. B}
  \bibinfo{volume}{\textbf{54}}(13), \bibinfo{pages}{8992} (\bibinfo{date}{Oct.
  1996}).
\bibitem{Hytch1995}
\bibinfo{author}{M.~J. H\"ytch} and \bibinfo{author}{M.~Gandais},
  \bibinfo{journal}{Philosophical Magazine A} \bibinfo{volume}{\textbf{72}}(3),
  \bibinfo{pages}{619} (\bibinfo{date}{1995}).
\bibitem{Gong1996}
\bibinfo{author}{G.~Q. Gong}, \bibinfo{author}{A.~Gupta},
  \bibinfo{author}{G.~Xiao}, \bibinfo{author}{P.~Lecoeur}, and
  \bibinfo{author}{T.~R. McGuire}, \bibinfo{journal}{Phys. Rev. B}
  \bibinfo{volume}{\textbf{54}}(6), \bibinfo{pages}{R3742} (\bibinfo{date}{Aug.
  1996}).
\bibitem{Jo1999}
\bibinfo{author}{M.-H. Jo}, \bibinfo{author}{N.~D. Mathur},
  \bibinfo{author}{J.~E. Evetts}, \bibinfo{author}{M.~G. Blamire},
  \bibinfo{author}{M.~Bibes}, and \bibinfo{author}{J.~Fontcuberta},
  \bibinfo{journal}{Applied Physics Letters} \bibinfo{volume}{\textbf{75}}(23),
  \bibinfo{pages}{3689} (\bibinfo{date}{1999}).
\bibitem{Gonzalez2008}
\bibinfo{author}{I.~Gonzalez}, \bibinfo{author}{S.~Okamoto},
  \bibinfo{author}{S.~Yunoki}, \bibinfo{author}{A.~Moreo}, and
  \bibinfo{author}{E.~Dagotto}, \bibinfo{journal}{Journal of Physics: Condensed
  Matter} \bibinfo{volume}{\textbf{20}}(26), \bibinfo{pages}{264002 (7pp)}
  (\bibinfo{date}{2008}).
\bibitem{Adamo2009}
\bibinfo{author}{C.~Adamo}, \bibinfo{author}{C.~A. Perroni},
  \bibinfo{author}{V.~Cataudella}, \bibinfo{author}{G.~D. Filippis},
  \bibinfo{author}{P.~Orgiani}, and \bibinfo{author}{L.~Maritato},
  \bibinfo{journal}{Physical Review B (Condensed Matter and Materials Physics)}
  \bibinfo{volume}{\textbf{79}}(4), \bibinfo{pages}{045125},
  \bibinfo{eid}{045125} (\bibinfo{numpages}{5}  pages)  (\bibinfo{date}{2009}).
\bibitem{Yahia2003}
\bibinfo{author}{M.~Yahia} and \bibinfo{author}{H.~Batis},
  \bibinfo{journal}{European Journal of Inorganic Chemistry}
  \bibinfo{volume}{\textbf{2003}}(13), \bibinfo{pages}{2486}
  (\bibinfo{date}{2003}).

\end{thebibliography}
\end{document}